\documentclass[12pt]{article}
\usepackage{amsmath,amssymb,amsfonts,color,graphicx,cite,color,feynarts}
\usepackage{latexsym}
\usepackage[normalem]{ulem}
\usepackage{epsfig,psfrag,rotating,soul}
\usepackage{rotfloat}
\input paperdef

\evensidemargin \oddsidemargin
\allowdisplaybreaks

\begin{document}
\thispagestyle{empty}

\def\thefootnote{\fnsymbol{footnote}}

\begin{flushright}
\mbox{}
IFT--UAM/CSIC--22-129 
\end{flushright}

\vspace{0.5cm}

\begin{center}

\begin{large}
\textbf{Non-Holomorphic Soft-Term Contributions to the Higgs-Boson}
\\[2ex]
\textbf{Masses in the Feynman Diagrammatic Approach}
\end{large}

\vspace{1cm}

{\sc
M.~Rehman$^{1}$%
\footnote{email: m.rehman@comsats.edu.pk}%
, S.~Heinemeyer$^{2}$%
\footnote{email: Sven.Heinemeyer@cern.ch}%
}

\vspace*{.7cm}
{\sl
${}^1$Department of Physics, Comsats University Islamabad, 44000
  Islamabad, Pakistan \\[.1em] 
${}^2$Instituto de F\'isica Te\'orica, (UAM/CSIC), Universidad
  Aut\'onoma de Madrid, Cantoblanco, 28049 Madrid, Spain
}
\end{center}

\vspace*{0.1cm}

\begin{abstract}
\noindent

We study the effects of non-holomprphic soft SUSY-breaking terms added
to the Minimal Supersymmetric Standard Model (MSSM) on the Higgs-boson
masses. The calculation of the non-holomorphic contributions is perfromed
at the one-loop level in the Feynman diagramatic approach.
After generating a \fa\ model file with the help of {\tt SARAH}, we
calculate the renormalized Higgs-boson self-energies at the one-loop level using the
\fa/\fc\ setup. The results obtained from \fa/\fc\ are fed to \fh\
to estimate the contributions to the neutral \cp-even Higgs-boson
masses, $M_{h,H}$, as well as to the charged Higgs-boson mass, $\MHp$.
For the specific set of
parameter points that we choose for this study, the non-holomorphic
soft-term contributions to the \cp-even light Higgs boson mass
$M_h$, contrary to claims in the literature, turn out to be
very small. For the \cp-even heavy Higgs boson mass,
$M_{H}$, as well as for the charged Higgs boson mass, $\MHp$,
the contributions can be substantially larger for some parts of
parameter space.

\end{abstract}

\def\thefootnote{\arabic{footnote}}
\setcounter{page}{0}
\setcounter{footnote}{0}

\newpage


\section{Introduction}
\label{sec:intro}

After the Higgs discovery at the Large Hadron Collider
(LHC)~\cite{Aad:2012tfa,Chatrchyan:2012xdj}, much attention has been
devoted to measure the mass and other properties of the Higgs boson as 
precisely as possible. These properties, within the current experimental
and theoretical uncertainties, are in agreement with the Standard Model
(SM) predictions~\cite{Sekmen:2022vzu}. Consequently, any model beyond the
SM (BSM) should contain a state matching the LHC mass and rate
measurements. The requirement imposes strong
constraints on any BSM parameter space. On the other hand, the absense
of a discovery of any (additional) BSM particles also impose severe
constraints on the new physics parameters. Due to these reasons, Minimal
Supersymmetric Standard Model (MSSM)~\cite{mssm}, which is probably the
best and most studied extension of the SM, is also facing severe
constraints on its parameter space. For example, the present value of
the Higgs boson mass $M^{\rm obs}_{H}=125.38 \pm 0.14\gev$~\cite{CMS:2020xrn}
requires the SUSY partners of the top quark, the
scalar tops, to be either in the multi-TeV range, or that some 
relations among their parameters are fulfilled. One way to deal with this
problem would be to look for extra sources to the radiative corrections
while keeping the sfermion masses light. Left-right mixing and flavor
mixing in the sfermions can serve the purpose to some
extent~\cite{AranaCatania:2011ak,Gomez:2014uha}. See
\citere{Slavich:2020zjv} for a recent review on SUSY Higgs-boson mass
calculations. 

In this paper we will take a different route. In the MSSM, the
superpotential and soft breaking terms are generally considered as
holomorphic functions. While the superpotential must be holomorphic, the
soft SUSY-breaking terms can be non-holomorphic (NH) in
nature~\cite{Girardello:1981wz, Bagger:1993ji}. For further reference,
we will call this setup the Non-Holomorphic Supersymmetric Standard
Model (NHSSM). 
In general, the NH soft SUSY-breaking terms 
can contribute to the radiative corrections to
Higgs boson masses.
The question arises whether these additional freedom in
the form of NH soft SUSY-breaking terms have the potential to increase
the value of the light \cp-even Higgs boson in a relevant way. 
Some of the previous analyses of the NH effects
particularly to the Higgs boson sector can be found in
\citeres{Jack:1999ud, Jack:1999fa, Jack:2004dv, Cakir:2005hd,
  Sabanci:2008qp, Un:2014afa}.  More recently in
\citere{Chattopadhyay:2016ivr}, it was reported that the NH soft
SUSY-breaking terms can enhance/decrease the light \cp-even Higgs-boson
mass $M_h$ by up to $3\gev$. The analyses in this regard focused only on
the leading top-stop contibutions to the Higgs sector. In
\citere{Chattopadhyay:2016ivr} the effects of NH soft SUSY-breaking
terms on $M_h$  via their effects on the scalar top masses were
calculated. The NH terms enter the scalar top sector through left-right
mixing parameter $X_t$. However, the observed effects can possibly be
mimicked by a change in the holomorphic soft SUSY-breaking terms, in
particular the trilinear Higgs-stop coupling, $A_t$: for each choice of
$A_t^\prime$ the parameter $A_t$ can be adjusted to yield the same
scalar top masses and mixing angles. An observed scalar top mass
spectrum thus corresponds 
to a continuous set of combinations of $A_t$ and $A_t^\prime$ (keeping
the other soft SUSY-breaking parameters and $\mu$ fix). An analysis that
simply varies $A_t^\prime$, resulting in shifts in the scalar top
masses and mixing angle, can thus not be regarded realistic. On
the other hand, 
the analysis in \citere{Chattopadhyay:2016ivr} neglected the effects of
the NH terms entering the 
Higgs-sfermion couplings. As NH soft SUSY-breaking terms also enter in
the couplings of the Higgs bosons to the scalar fermions, it is
important to consider all possible effects simultanously, while clearly
working out the genuine NH effects. 

In this work, we have calculated the effects of
NH soft SUSY-breaking term $\At^\prime$ to the Higgs boson mass
spectra at one-loop 
level using the Feynman diagramatic approach. These newly evaluated
one-loop corrections, obtained by
\fa/\fc\ setup~\cite{Hahn:2000kx,Hahn:2001rv,Fritzsche:2013fta,Hahn:1998yk},
were then fed into
\fh~\cite{mhiggslong, mhiggsAEC, mhcMSSMlong, Frank:2006yh, Mh-logresum,
  Bahl:2016brp, Bahl:2017aev, Bahl:2018qog} such that
all other known higher-order corrections can be taken over from the
MSSM. We ensured that the stop spectrum does not change under the
  variation of $\At^\prime$.
This allows to reliably estimate the effects of NH soft
SUSY-breaking terms to the Higgs boson masses at the one-loop level.

The paper is organized as follows: first we present the main features of
the NHSSM in \refse{sec:model_NHSSM}. The computational setup is given
in \refse{sec:CompSetup}. The numerical results are presented in
\refse{sec:NResults}. Our conclusions can be found in
\refse{sec:conclusions}.  

\section{Model set-up}
\label{sec:model_NHSSM}

The  MSSM is the simplest Supersymmetric structure one can build from the SM 
particle content. The general set-up for the soft SUSY-breaking
parameters is given by~\cite{mssm}
\begin{eqnarray}
\label{softbreaking}
-\cL_{\rm soft}&=&(m_{\tilde Q}^2)_i^j {\tilde q}_{L}^{\dagger i}
{\tilde q}_{Lj}
+(m_{\tilde u}^2)^i_j {\tilde u}_{Ri}^* {\tilde u}_{R}^j
+(m_{\tilde d}^2)^i_j {\tilde d}_{Ri}^* {\tilde d}_{R}^j
\nonumber \\
& &+(m_{\tilde L}^2)_i^j {\tilde l}_{L}^{\dagger i}{\tilde l}_{Lj}
+(m_{\tilde e}^2)^i_j {\tilde e}_{Ri}^* {\tilde e}_{R}^j
\nonumber \\
& &+{\tilde m}^2_{1}h_1^{\dagger} h_1
+{\tilde m}^2_{2}h_2^{\dagger} h_2
+(B \mu h_1 h_2
+ {\rm h.c.})
\nonumber \\
& &+ ( A_d^{ij}h_1 {\tilde d}_{Ri}^*{\tilde q}_{Lj}
+A_u^{ij}h_2 {\tilde u}_{Ri}^*{\tilde q}_{Lj}
+A_l^{ij}h_1 {\tilde e}_{Ri}^*{\tilde l}_{Lj}
\nonumber \\
& & +\frac{1}{2}M_1 {\tilde B}_L^0 {\tilde B}_L^0
+\frac{1}{2}M_2 {\tilde W}_L^a {\tilde W}_L^a
+\frac{1}{2}M_3 {\tilde G}^a {\tilde G}^a + {\rm h.c.}).
\end{eqnarray}
Here $m_{\tilde Q}^2$ and $m_{\tilde L}^2$ are $3 \times 3$
matrices in family space (with $i,j$ being the
generation indices) for the soft SUSY-breaking masses of the
left handed squark ${\tilde q}_{L}$ and slepton ${\tilde l}_{L}$
$SU(2)$ doublets, respectively. $m_{\tilde u}^2$, $m_{\tilde d}^2$ and
$m_{\tilde e}^2$ contain the soft masses for right handed up-type squark
${\tilde u}_{R}$,  down-type squarks ${\tilde d}_{R}$ and charged
slepton ${\tilde e}_{R}$ $SU(2)$ singlets, respectively. $A_u$, $A_d$
and $A_l$ are the $3 \times 3$ matrices for the trilinear
couplings for up-type squarks, down-type 
squarks and charged sleptons, respectively. 
$\mu$ is Higgs mixing paramter, ${\tilde m}_1$, ${\tilde m}_2$
and $B$ are the soft 
SUSY-breaking parameters of the Higgs sector, where $h_1$ and $h_2$
denote the two doublets.
In the last line $M_1$, $M_2$ and $M_3$
define the bino, wino  and gluino mass terms, respectively.

The superpotential in the MSSM must be holomorphic, and consequently,
the soft SUSY-breaking sector is generally paramterized via holomorphic
operators. However the MSSM can be extended by introducing 
R-Parity violating and/or non-holomorphic terms in the soft breaking
sector~\cite{Girardello:1981wz, Bagger:1993ji, Chakrabortty:2011zz}. 
In it's simplest form the following terms can be introduced in
the soft SUSY-breaking sector of the MSSM: 
\begin{eqnarray}
\label{NonH-TrilinearTerms}
-\cL_{\rm soft}^{\rm NH}&=&A_d^{^\prime ij}h_2 {\tilde d}_{Ri}^*{\tilde q}_{Lj}
+A_u^{^\prime ij}h_1 {\tilde u}_{Ri}^*{\tilde q}_{Lj}
+A_l^{^\prime ij}h_2 {\tilde e}_{Ri}^*{\tilde l}_{Lj}
+\mu^{\prime} {\tilde h}_1 {\tilde h}_2
\end{eqnarray}
Here $\mu^{\prime}$ is NH Higgsino mass term, whereas $A_u^{^\prime}$,
$A_d^{^\prime}$ and $A_l^{^\prime}$ denote the NH trilinear
coupling matrices for up-type squarks, down-type squarks and charged
sleptons, respectively. These terms do not necessarily have any
relationship with the holomorphic trilinear soft terms given in
\refeq{softbreaking}.
One possibility is to assume them equal to the holomorphic
trilinear couplings as a ``boundary condition'' at the GUT scale in
models such as the Constrained MSSM. 
However, even in that case renormalization group
equation running effects will result in completly different
non-holomorphic trinlinear terms\cite{Un:2014afa}. Therefore it is a
sensible approach to consider non-holomorphic trilinear terms
independent, but overall of the same order of magnitude as the usual
trinlinear couplings while comparing the NHSSM predictions with the
experimental results.   

In the presence of the non-holomorphic trilinear terms, the sfermion
mass matrices will be modified as  
\begin{equation}
M_{\tilde{f}}^{2}=\left(
\begin{array}
[c]{cc}%
m_{\tilde{f}LL}^{2} & m_{\tilde{f}LR}^{2}\\[.5em]
m_{\tilde{f}LR}^{2\dag} & m_{\tilde{f}^{\prime}RR}^{2}%
\end{array}
\right) \label{fermion mass matrix}%
\end{equation}
with%
\begin{align}
m_{\tilde{f}LL}^{2}  & =m_{\tilde{f}}^{2}+M_{Z}^{2}\cos2\beta\left(  I_{3}%
^{f}-Q_{f}s_{W}^{2}\right)  +m_{f}^{2}\nonumber\\
m_{\tilde{f}^{\prime}RR}^{2}  & =m_{\tilde{f}^{%
\acute{}%
}}^{2}+M_{Z}^{2}\cos2\beta Q_{f^{\prime}}s_{W}^{2}+m_{f}^{2}\nonumber\\
m_{\tilde{f}LR}^{2}  & =m_{f}X_{f}\text{ ;
\ \ \ }X_{f}=A_{f}-(\mu+A_f^\prime)\left\{  \cot\beta;\tan\beta\right\} \label{mass terms}%
\end{align}
where $I_{3}^{f}$ is the weak isospin of fermions, $Q_{f}$ is the EM charge,
$m_{f}$ is the standard fermion mass, $f$ and $f^{\prime}$ stands for
left and right handed sfermions (except for neutrino),
respectively.
$\MZ$ and $\MW$ denote the mass of the $Z$~and the $W$~boson, and
$s_W = \sqrt{1 - c_W^2}$ with $c_W = \MW/\MZ$.
$A_{f}$ ($A^{\prime}_{f}$) 
is the holomorphic (non-holomorphic) trilinear coupling\footnote{
We neglect \cp~ violation throughout the paper.},
$\mu$ is the Higgs mixing parameter, and $\cot\beta$ is for up type
squarks and $\tan\beta$ is for down type squarks and charged
sleptons
($\tan \beta := v_2/v_1$, is the ratio of the two vacuum
expectation values of the two Higgs doublets.)
The NH higgsino mass parameter $\mu^{\prime}$ mentioned in
\refeq{NonH-TrilinearTerms} modifies the neutralino and chargino mass
matrices, but it does not enter into the modified sfermion mass
matrix. Consequently, it will not be particularly relavent for our present
analysis as we focus on the top/stop contributions.   
It should be noted that because of the different combination of fields
in $\cL_{\rm soft}^{\rm NH}$ w.r.t.\ $\cL_{\rm soft}$ the
non-holomorphic trilinear couplings $A_f^\prime$ receive the additional
factors of $\tb$ or $\CTb$. 

As discussed before, the the NH trilinear terms also modify the
Higgs-sfermion-sfermion couplings. Here we show the couplings of the
lightest Higgs boson $h$ to up-type squarks.
\begin{align}
C(h,\tilde{u}_{i}^{s},\tilde{u}_{j}^{t}) &  =\frac{-ie\delta_{ij}}{6M_{W}%
c_{W}s_{W}s_{\beta}}\Big[3c_{W}m_{u_{i}}\{A_{ii}^{u}c_{\alpha}+(\mu
+A_{ii}^{\prime u})s_{\alpha}\}U_{s,1}^{\tilde{u},i}U_{t,2}^{\tilde{u}%
,i}\nonumber\\
&  +\{6c_{\alpha}c_{W}m_{u_{i}}^{2}-M_{W}M_{Z}s_{\alpha+\beta}s_{\beta
}(3-4s_{W}^{2})\}U_{s,1}^{\tilde{u},i}U_{t,1}^{\tilde{u},i}\nonumber\\
&  +\{6c_{\alpha}c_{W}m_{u_{i}}^{2}-4M_{W}M_{Z}s_{\alpha+\beta}s_{\beta}%
s_{W}^{2}\}U_{s,2}^{\tilde{u},i}U_{t,2}^{\tilde{u},i}\nonumber\\
&  +3c_{W}m_{u_{i}}\{A_{ii}^{u}c_{\alpha}+(\mu+A_{ii}^{\prime u})s_{\alpha
}\}U_{s,2}^{\tilde{u},i}U_{t,1}^{\tilde{u},i}\Big]\label{ChSqSq}%
\end{align}
The coupling of the charged Higgs boson $H^{-}$ to up-type and
down-type squarks is given by 
\begin{align}
C(H^{-},\tilde{u}_{i}^{s},\tilde{d}_{j}^{t}) &  =\frac{ieV_{ij}^{CKM}}%
{2M_{W}s_{W}s_{\beta}}\Big[m_{u_{i}}U_{s,2}^{\tilde{u},i}U_{t,1}^{\tilde{d}%
,j}\{A_{ii}^{u}+(\mu+A_{ii}^{\prime u})t_{\beta}\}\nonumber\\
&  +m_{u_{i}}m_{d_{j}}U_{s,2}^{\tilde{u},i}U_{t,2}^{\tilde{d},j}(1+t_{\beta
}^{2})+U_{s,1}^{\tilde{u},i}U_{t,2}^{\tilde{d},j}mt_{\beta}\{A_{ii}%
^{d}t_{\beta}+(\mu+A_{ii}^{\prime d})\}\nonumber\\
&  +U_{s,1}^{\tilde{u},i}U_{t,1}^{\tilde{d},j}\{m_{u_{i}}^{2}-t_{\beta}%
(M_{W}^{2}s_{2\beta}-m_{d_{j}}^{2})t_{\beta}\}\Big]
\label{CHpSqSq}
\end{align}
Here $i,j$ are the generation indices (we assume flavor
conservation throughout the paper), $U_{s,s'}^{\tilde{u},i}$
($U_{t,t'}^{\tilde{d},j}$) is the 
$2\times 2$ rotation matrix for up-type (down-type)
squarks, and we use the short hand notation $s_x, c_x, t_x$ for
$\sin x$, $\cos x$, $\tan x$, respectively,
where $\alpha$ is the \cp-even Higgs mixing angle.
The couplings of the \cp-even heavy Higgs boson $H$
to the up-type squarks can be obtained by replacing $c_{\alpha}
\rightarrow s_{\alpha}$, $s_{\alpha} \rightarrow -c_{\alpha}$ and
$s_{\alpha+\beta} \rightarrow -c_{\alpha+\beta}$ in \refeq{ChSqSq}. 
It is interesting to observe that the $A_t^\prime$ enter differently
into the scalar top masses and into the trilinear Higgs-stop
couplings. This will be crucial for our numerical analysis, see the
discussion in \refse{sec:NResults}. 

\section{Higher order corrections in the NHSSM Higgs sector}
\label{sec:CompSetup}

\subsection{Tree-level structure and higher-order corrections}

The MSSM (and thus the NHSSM) Higgs-boson sector consist of two Higgs
doublets and predicts the existence of five physical
Higgs bosons, the light and heavy $\cp$-even $h$ and $H$, the $\cp$-odd $A$,
and a pair of charged Higgs bosons, $H^\pm$. At the tree-level the Higgs
sector is described with the help of two parameters: the mass of the
$A$~boson, $\MA$, and $\tb = v_2/v_1$, the ratio of the two vacuum
expectation values. 
The tree-level relations and in particular the tree-level masses receive
large higher-order corrections, see, 
e.g., \citeres{MHreviews, Draper:2016pys,Slavich:2020zjv} and references
therein. 

The lightest MSSM Higgs boson, with mass $\Mh$, can be interpreted as
the new state discovered at the LHC around
$\sim 125 \gev$~\cite{Heinemeyer:2011aa}. 
The present experimental uncertainty at the LHC for $\Mh$, 
is about~\cite{CMS:2020xrn},
\begin{align}
\de\Mh^{\rm exp,today} \sim 140 \mev~.
\end{align}
This can possibly be reduced below the level of 
\begin{align}
\de\Mh^{\rm exp,future} \lsim 50 \mev
\end{align}
at future $e^+e^-$~colliders~\cite{dbd}.
Similarly, for the masses of the heavy neutral Higgs 
$\MH$, an uncertainty at the $1\%$ level 
could be expected at the LHC~\cite{cmsHiggs}.

\medskip
In the Feynman diagrammatic (FD) approach that we are following in our
calculation here, the higher-order corrected  $\cp$-even Higgs boson
masses are obtained by finding the 
poles of the $(h,H)$-propagator 
matrix. The inverse of this matrix is given by
\BE
\left(\Delta_{\rm Higgs}\right)^{-1}
= - i \ML p^2 -  \mHtree^2 + \hSi_{HH}(p^2) &  \hSi_{hH}(p^2) \\
     \hSi_{hH}(p^2) & p^2 -  \mhtree^2 + \hSi_{hh}(p^2) \MR~.
\label{higgsmassmatrixnondiag}
\end{equation}
Determining the poles of the matrix $\Delta_{\rm Higgs}$ in
\refeq{higgsmassmatrixnondiag} is equivalent to solving
the equation
\begin{equation}
\left[p^2 - \mhtree^2 + \hSi_{hh}(p^2) \right]
\left[p^2 - \mHtree^2 + \hSi_{HH}(p^2) \right] -
\left[\hSi_{hH}(p^2)\right]^2 = 0\,.
\label{eq:proppole}
\end{equation}

Similarly, in the case of the charged Higgs sector, the corrected Higgs
mass is derived  by the position of the pole in the charged Higgs
propagator (for details please see \citere{Frank:2013hba} and references
therein), which is defined by:   
\noindent \begin{equation}
p^{2}-m^{2}_{H^{\pm},{\rm tree}} +
\hat{\Sigma}_{H^{-}H^{+}}\left(p^{2}\right)=0.
\label{eq:proppolech}
\end{equation}

The (renormalized) Higgs-boson self-energies in \refeqs{eq:proppole} and
\ref{eq:proppolech} can be evaluated at the $n$-loop 
level, by an explicit (FD) calculation of the corresponding loop diagrams. 
As discussed above, in this work we will concentrate on the one-loop
corrections from the top/stop sector. The FD contributions to the
Higgs-boson self-energies can be supplemented by a resummation of
leading and subleading logarithmic contributions, which are relevant in
the case of heavy scalar tops. For more details, see
\citere{Slavich:2020zjv}. This will be relevant for the numerical
evaluation presented below in \refse{sec:NResults}. 


\subsection{Non-holomorhpic Contributions to Higgs Sector}
\label{sec:strategy0}

\begin{figure}[htb!]
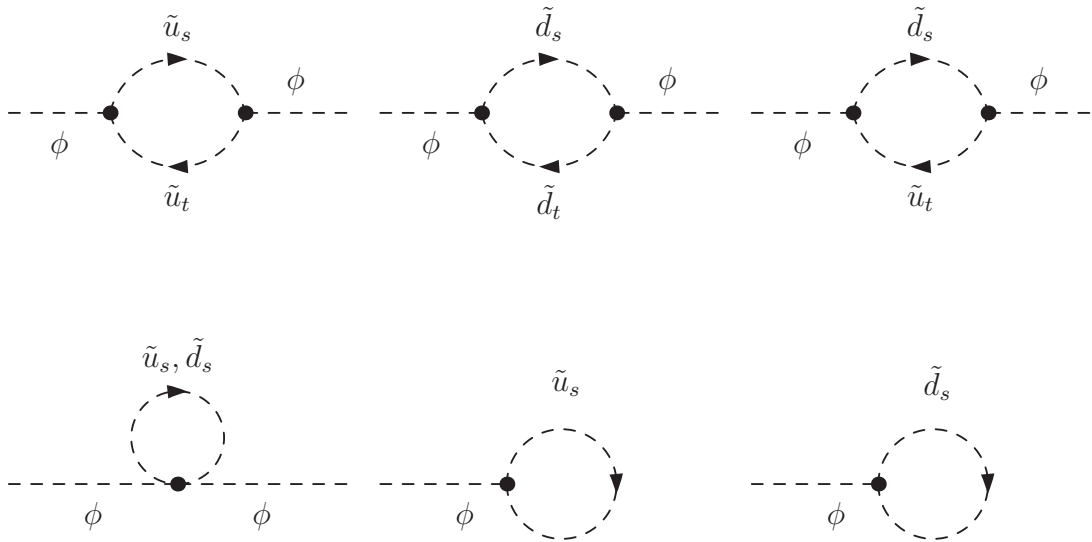

\begin{center}

\unitlength=1.0bp%

\begin{feynartspicture}(432,280)(3,2)

\FADiagram{}
\FAProp(0.,10.)(6.,10.)(0.,){/ScalarDash}{0}
\FALabel(3.,8.93)[t]{$\phi$}
\FAProp(20.,10.)(14.,10.)(0.,){/ScalarDash}{0}
\FALabel(17.,11.07)[b]{$\phi$}
\FAProp(6.,10.)(14.,10.)(0.8,){/ScalarDash}{-1}
\FALabel(10.,5.73)[t]{$\tilde u_{t}$}
\FAProp(6.,10.)(14.,10.)(-0.8,){/ScalarDash}{1}
\FALabel(10.,14.27)[b]{$\tilde u_{s}$}
\FAVert(6.,10.){0}
\FAVert(14.,10.){0}

\FADiagram{}
\FAProp(0.,10.)(6.,10.)(0.,){/ScalarDash}{0}
\FALabel(3.,8.93)[t]{$\phi$}
\FAProp(20.,10.)(14.,10.)(0.,){/ScalarDash}{0}
\FALabel(17.,11.07)[b]{$\phi$}
\FAProp(6.,10.)(14.,10.)(0.8,){/ScalarDash}{-1}
\FALabel(10.,5.73)[t]{$\tilde d_{t}$}
\FAProp(6.,10.)(14.,10.)(-0.8,){/ScalarDash}{1}
\FALabel(10.,14.27)[b]{$\tilde d_{s}$}
\FAVert(6.,10.){0}
\FAVert(14.,10.){0}

\FADiagram{}
\FAProp(0.,10.)(6.,10.)(0.,){/ScalarDash}{0}
\FALabel(3.,8.93)[t]{$\phi$}
\FAProp(20.,10.)(14.,10.)(0.,){/ScalarDash}{0}
\FALabel(17.,11.07)[b]{$\phi$}
\FAProp(6.,10.)(14.,10.)(0.8,){/ScalarDash}{-1}
\FALabel(10.,5.73)[t]{$\tilde u_{t}$}
\FAProp(6.,10.)(14.,10.)(-0.8,){/ScalarDash}{1}
\FALabel(10.,14.27)[b]{$\tilde d_{s}$}
\FAVert(6.,10.){0}
\FAVert(14.,10.){0}

\FADiagram{}
\FAProp(0.,10.)(10.,10.)(0.,){/ScalarDash}{0}
\FALabel(5.,8.93)[t]{$\phi$}
\FAProp(20.,10.)(10.,10.)(0.,){/ScalarDash}{0}
\FALabel(15.,8.93)[t]{$\phi$}
\FAProp(10.,10.)(10.,10.)(10.,15.5){/ScalarDash}{-1}
\FALabel(10.,16.57)[b]{$\tilde u_{s}, \tilde d_{s}$}
\FAVert(10.,10.){0}

\FADiagram{}
\FAProp(0.,10.)(7.5,10.)(0.,){/ScalarDash}{0}
\FALabel(5.,8.93)[t]{$\phi$}
\FAProp(7.5,10.)(7.5,10.)(14.,10.){/ScalarDash}{-1}
\FALabel(11.,16.)[]{$\tilde u_{s}$}
\FAVert(7.5,10.){0}

\FADiagram{}
\FAProp(0.,10.)(7.5,10.)(0.,){/ScalarDash}{0}
\FALabel(5.,8.93)[t]{$\phi$}
\FAProp(7.5,10.)(7.5,10.)(14.,10.){/ScalarDash}{-1}
\FALabel(11.,16.)[]{$\tilde d_{s}$}
\FAVert(7.5,10.){0}

\end{feynartspicture}
\end{center}
\caption{
Generic Feynman diagrams for the Higgs boson self-energies and
tadpoles. $\phi$ denotes any of the Higgs bosons, $h$, $H$, $A$ or 
$H^\pm$; $u$ stand for $u,c,t$; $d$ stand for $d,s,b$; $\tilde u_{s,t}$
and $\tilde d_{s,t}$ are the six mass 
eigenstates of up-type and down-type squarks, respectively.} 
\label{FeynDiagHSelf}
\end{figure}

The NH soft SUSY-breaking parameters enter into the one-loop
prediction of the various (renormalized) Higgs-boson self-energies and
tadpoles. As discussed above, they can enter into the scalar fermion
masses, where, however, their effect can be compensated by a change in
the corresponding holomorphic trilinear coupling. They also inter into
the Higgs-sfermion-sfermion coupling, see \refeq{ChSqSq}, which will
have the main effect in our analysis. 
Generic Feynman diagrams that involve non holomorphic
couplings are shown in the \reffi{FeynDiagHSelf}. Here we restrict
ourself to quark/squark contributions only.

In the following, we briefly describe our work flow  for the
calculation.
To calculate the non holomprphic contributions to the
Higgs boson self energies, we first created NHSSM model file for \fa\
using Mathematica package SARAH~\cite{Staub:2009bi, Staub:2010jh,
Staub:2012pb, Staub:2013tta, Staub:2015kfa}.
The
\fa/\fc~\cite{Hahn:2000kx,Hahn:2001rv,Fritzsche:2013fta,Hahn:1998yk}
packages have then be used to analytically calculate the NHSSM
contributions to the Higgs boson self-energies, given as a
function of $\At$ and $\At^\prime$.
For the numerical evaluation with the \fa/\fc\ setup the \fc\ driver
files had to be adjusted from the MSSM to the NHSSM.
Concerning the numerical evaluation, for a given value of
$\At^{\rm MSSM}$ in the MSSM and $\At^\prime$ in the 
NHSSM a new value of $\At^{\rm NHSSM}$ is calculated such that
$\Xt^{\rm MSSM} = \At^{\rm MSSM} - \mu\CTb$ and
$\Xt^{\rm NHSSM} = \At^{\rm NHSSM} - (\mu + \At^\prime)\CTb$ are
identical (yielding the same values for the stop masses and mixings,
see the discussion in the next section). Using $\At^{\rm NHSSM}$ and
$\At^\prime$ the NH contribution to the Higgs-boson self-energies is
calculated numerically. 
To avoid
double counting, we subtracted the Higgs-boson self-energy values at
$A_t^{\prime}=0$ (i.e.\ $\At^{\rm MSSM} \equiv \At^{\rm NHSSM}$)
from the obtained results.

These numerical values were fed to 
\fh~\cite{feynhiggs,mhiggslong,mhiggsAEC,mhcMSSMlong,Mh-logresum,Bahl:2016brp,
  Bahl:2017aev,Bahl:2018qog} using the \fh\ function
{\tt FHAddSelf} (where in \fh\ the value $\At^{\rm MSSM}$ was used).
The \fh\ package already contains the complete set of
one-loop corrections in the MSSM. Those are supplemented with leading
and sub-leading two-loop corrections as well as a resummation of leading
and sub-leading logarithmic contributions from the $t/\Stop$ sector.
In this way we include the NH contributions from $\At^\prime$
into the most precise evaluation of the MSSM Higgs-boson masses
available. This allows 
us to readily estimate the effect of the NH soft SUSY-breaking terms.


\section{Numerical Results}
\label{sec:NResults}

\subsection{General strategy}
\label{sec:strategy}

The leading corrections to $M_h$ from the top/scalar top loops in the
NHSSM have been calculated in \citere{Chattopadhyay:2016ivr} and are given by 
\begin{equation}
\Delta m^{2}_{h,t/\Stop}=\frac{3 g_2^2 m^4_t}{8 \pi^2 M^2_W }   \left[\ln \left(\frac{m_{\tilde t_{1}} m_{\tilde t_{2} }}{m^2_{t}}\right)+\frac{X_t^2}{m_{\tilde t_{1}} m_{\tilde t_{2} }}\left(1-\frac{X_t^2}{12 m_{\tilde t_1} m_{\tilde t_2}}\right)\right]
\label{deltaMh-NHSSM}
\end{equation}
where $\Xt^{\rm NHSSM} =: X_t=A_t-(\mu+A_t^{\prime})\CTb$.
The non-holomorphic 
trilinear coupling $A_t^{\prime}$ affects the $X_t$ parameter as well as
the scalar top quark masses and mixing angle. A simple change in
the value of 
$A_t^{\prime}$ with fix $A_t$ will result in the change of $X_t$ which
in turn will change $M_h$. However, with this approach, we can not
distinguish the pure NHSSM contribution, as the same results can be
obtained by a correspondingly changed value of $A_t$ in the MSSM.
Moreover, (if SUSY is realized in nature) the scalar top masses
and mixing will
be known in the future, and the choice of the soft SUSY-breaking
parameters have to reproduce their values. Therefore it makes sense to
analyze the NH effects in a scheme that allows to keep the two stop
masses and the mixing angle fixed. 
On the other hand, in the FD approach the $A_t^{\prime}$ appears also in
the coupling 
of the Higgs boson to the scalar top quarks. In order to estimate the
contribtuions on the Higgs-boson mass spectrum coming purely from
$A_t^{\prime}$, it is therefore
important to fix the value of $X_t$ parameter (see the discussion
in the previous section), shifting the NH effects
completely into the change in the Higgs-stop coupling. 


\subsection{Benchmark scenarios}
\label{sec:bench}

In our numerical analyses we have followed the above described approach.
We evaluated the results in three benchmark scenarios defined
in \citere{Bahl:2018zmf} that are used by the
ATLAS and CMS collaboration for their interpretation of MSSM Higgs boson
searches.
These are the $M_h^{125}$ scenario (heavy SUSY particles, effectively
the Two Higgs Doublet Model type~II with SUSY restrictions on
Higgs-boson masses and couplings), the $M_h^{125}(\tilde\tau)$ scenario
(featuring light scalar taus) and the $M_h^{125}(\tilde\chi)$ scenrio
(featuring light charginos and neutralinos). 
The three scenarios are
compatible with the LHC searches for SUSY particles and yield a light
\cp-even Higgs boson with mass around 125 $\gev$ with SM-like
properties. For these scenarios indirect constraints like dark matter
density, flavor observables and the anomalous magnetic moment of the
muon on the MSSM parameters space are not taken into account on
purpose~\cite{Bahl:2018zmf}. These potential constraints mainly depend
on the parameters that are not important for Higgs-boson
phenomenology. Alternatively, small variations in the MSSM can
invalidate this type of constraints, while leaving the Higgs-boson
phenomenology largely unaffected, see the discussion in
\citere{Bahl:2018zmf}.
We furthermore assume that there is no (relevant) flavor
violation. Consequently, the first and second generation scalar fermions
have very mild effect on the predictions of the Higgs masses
and mixing. Thus a common soft SUSY-breaking mass $M_{\tilde{f}}=2\tev$
and corresponding Higg-sfermion interaction terms $A_f=0$ are assumed for
first and second generation sfermions in the benchmark scenarios. 
This is in full agreement with the
current exclusion bounds from CMS~\cite{CMS:2022goy} and
ATLAS~\cite{ATLAS:2020zms, ATLAS:2021twp}. In 
\refta{input-parameters}, we list the remaining soft SUSY-breaking input
paramters with corresponding scalar top masses for three scenarios
considered in our numerical analysis. 

\begin{table}[h!]
  \renewcommand{\arraystretch}{1.2}
\centerline{\begin{tabular}{|c||c|c|c|}
\hline
 & $M_h^{125}$ & $M_h^{125} (\tilde{\tau})$ & $M_h^{125} (\tilde{\chi})$ \\\hline
$m_{\tilde Q_{3}, \tilde U_{3}, \tilde D_{3}}$ & 1.5 $\tev$ & 1.5 $\tev$ & 1.5 $\tev$ \\
$m_{\tilde L_{3},\tilde E_{3}}$ & 2 $\tev$ & 350 $\gev$ & 2 $\tev$ \\
$\mu$ & 1 $\tev$ & 1 $\tev$ & 180 $\gev$ \\
$M_1$ & 1 $\tev$ & 180 $\gev$ & 160 $\gev$ \\
$M_2$ & 1 $\tev$ & 300 $\gev$ & 180 $\gev$ \\
$M_3$ & 2.5 $\tev$ & 2.5 $\tev$ & 2.5 $\tev$ \\
$X_t$ & 2.8 $\tev$ & 2.8 $\tev$ & 2.5 $\tev$ \\
$A_\tau$ & 0 & 800 $\gev$ & 0 \\
$A_b$ & 0 & 0 & 0 \\
\hline
$m_{\tilde t_{1}},m_{\tilde t_{2}}$ & 1339,1662  $\gev$ & 1339,1662
$\gev$ & 1358,1646 $\gev$ \\  
\hline
\end{tabular}}
\caption{Selected scenarios in the MSSM parameter space, taken from
  \citere{Bahl:2018zmf}.}
\label{input-parameters}
  \renewcommand{\arraystretch}{1.0}
\end{table}

For each scenario, we investigate three different combinations of $M_A$ and
$\tb$, taking into account the latest experimental limits for MSSM
Higgs-boson searches~\cite{ATLAS:2020zms,CMS:2022goy}:
\begin{itemize}
\item[P1]: $\MA = 1000 \gev$, $\tb = 7$
\item[P2]: $\MA = 1500 \gev$, $\tb = 15$
\item[P3]: $\MA = 2000 \gev$, $\tb = 45$
\end{itemize}
For our numerical analysis, the values of $A_t$ and $A_t^{\prime}$ have been
choosen such that the value of $X_t$ remain constant as given in the three
scenarios in \refta{input-parameters}.
However in order to extract pure NHSSM contributions we
treat $A_b$ and $A_{\tau}$ independent from $A_t$ (contrary to the
definition in \citere{Bahl:2018zmf}) and concentrate only
on top/stop sector.
Here it should be noted that the bottom/sbottom and tau/stau
contributions can also results in large radiative corrections to the
renormalized Higgs-boson self-energies due to the fact that the
corresponding non-holomorphic trilinear couplings 
$A_b^\prime$ and $A_{\tau}^\prime$ are multiplied by $\tb$.
However, a fixed value of $X_b (X_{\tau})$, as our strategy requires, 
can result in unrealistically large value of $A_b$
($A_{\tau}$). Furthermore, this can lead to severe numerical
instabilities in the evalution of the Higgs-boson spectra, even for moderate
values of $A_b^\prime$ and $A_{\tau}^\prime$, and special care
has to be taken to remain in a perturbative and numerically stable
regime of the model.
Consequently, here we
restrict ourselves to the corrections from the top/stop sector
(as it had been done in \citere{Chattopadhyay:2016ivr}), allowing
us to pin down the NH effects.
We leave a corresponding analysis of the effects of $\Ab^\prime$
and $\Atau^\prime$ for future work.


\subsection{NH contributions to renormalized Higgs-boson self-energies} 

In this subsection we present our results for the NH effects on the
renormalized Higgs-boson self-energies in the scenarios defined in the
previous subsection. 
To highlight the non-holomorphic contributions, we define
\begin{align}
\delta \hSi_{hh} &\eq \hSi_{hh} - \hSi_{hh}^{\rm MSSM}\,, \nonumber \\
\delta \hSi_{hH} &\eq \hSi_{hH} - \hSi_{hH}^{\rm MSSM}\,, \nonumber \\
\delta \hSi_{HH} &\eq \hSi_{HH} - \hSi_{HH}^{\rm MSSM}\,, \nonumber \\
\delta \hSi_{H^{\pm}} &\eq \hSi_{H^{\pm}} - \hSi_{H^{\pm}}^{\rm MSSM}\,,
\end{align}
and
\begin{align}
\delta \Mh &\eq \Mh - \Mh^{\rm MSSM}\,, \nonumber \\
\delta \MH &\eq \MH - \MH^{\rm MSSM}\,, \nonumber \\
\delta \MHp &\eq \MHp - \MHp^{\rm MSSM}\,,
\end{align}
where $\hSi_{hh}^{\rm MSSM}$, $\hSi_{hH}^{\rm MSSM}$, $\hSi_{HH}^{\rm MSSM}$,
$\hSi_{H^{\pm}}^{\rm MSSM}$, $\Mh^{\rm MSSM}$, $\MH^{\rm MSSM}$ and
$\MHp^{\rm MSSM}$ corresponds to the renormalized Higgs-boson self
energies and Higgs-boson masses with $A_t^{\prime} = 0$.   

\begin{figure}[ht!]
\vspace{0.5em}
  \begin{center}
\psfig{file=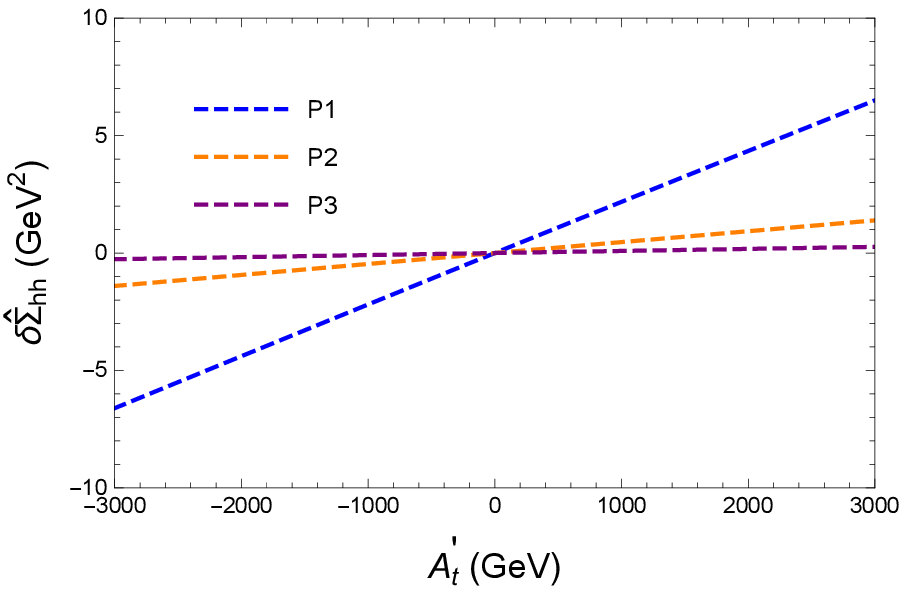  ,scale=0.75,angle=0,clip=}
\hspace{0.5cm}
\psfig{file=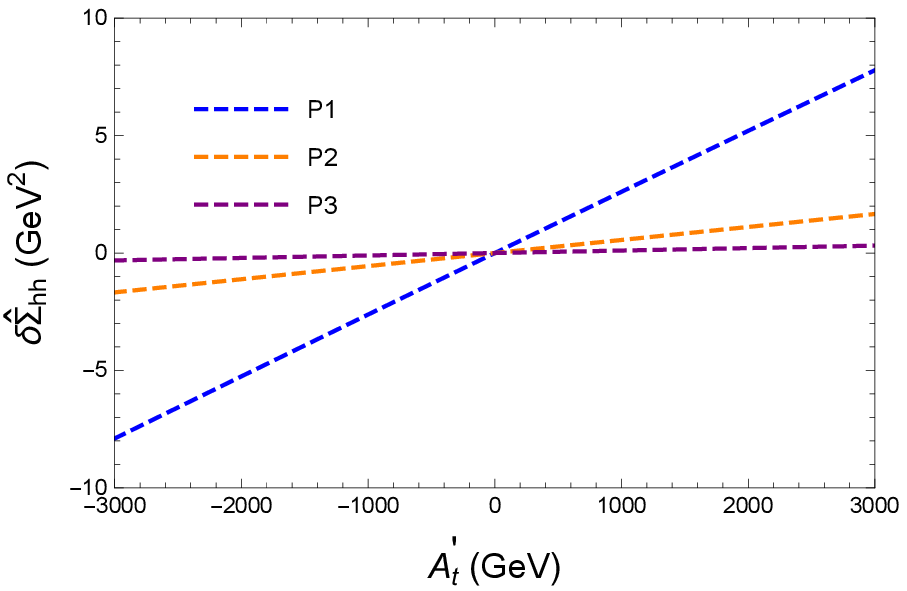 ,scale=0.75,angle=0,clip=}
\end{center}
\caption{ $\delta \hSi_{hh}$  as a funtion of $A^{\prime}_t$ for
  $M_h^{125}$ (left) and 
    $M_h^{125} (\tilde{\chi})$ (right plot). The results in
    $M_h^{125}(\tilde{\tau})$ are effectively identical to $M_h^{125}$
    and consequently not shown.}
\label{fig:SEh0}
\end{figure}

\begin{figure}[ht!]
\vspace{0.5em}
\begin{center}
\psfig{file=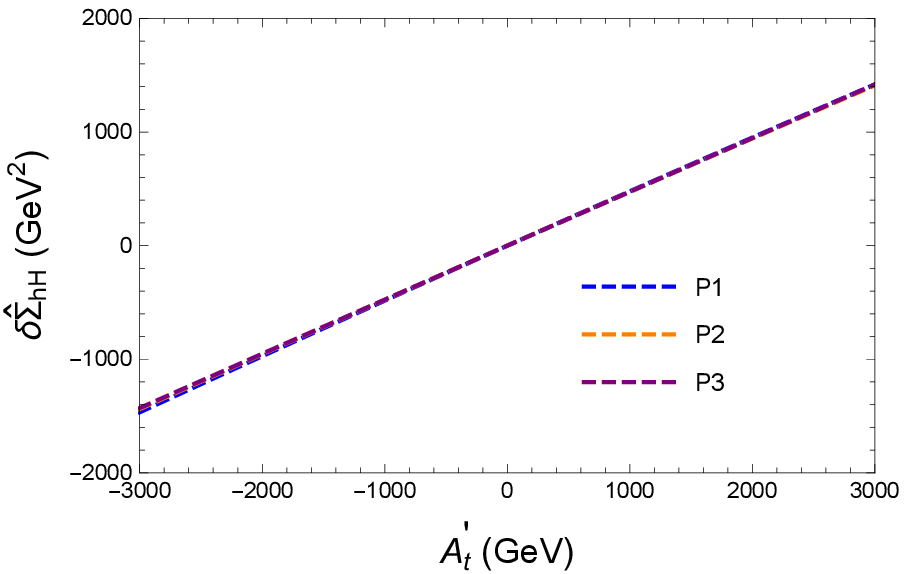  ,scale=0.75,angle=0,clip=}
\hspace{0.5cm}
\psfig{file=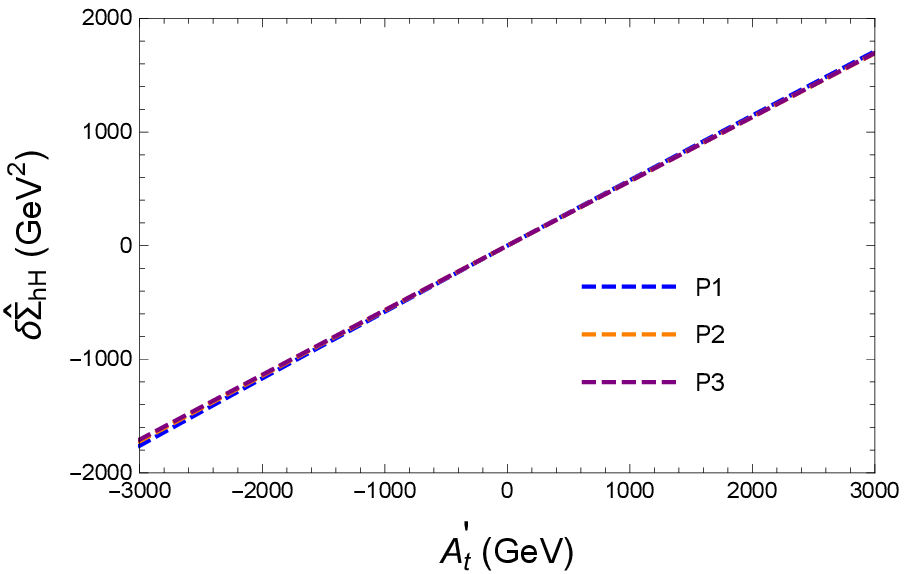 ,scale=0.75,angle=0,clip=}
\end{center}
\caption{ $\delta \hSi_{hH}$  as a funtion of $A^{\prime}_t$ for
  $M_h^{125}$ (left) and 
  $M_h^{125} (\tilde{\chi})$ (right plot).}    
\label{fig:SEh0HH}
\end{figure}

\begin{figure}[ht!]
\vspace{1em}
\begin{center}
\psfig{file=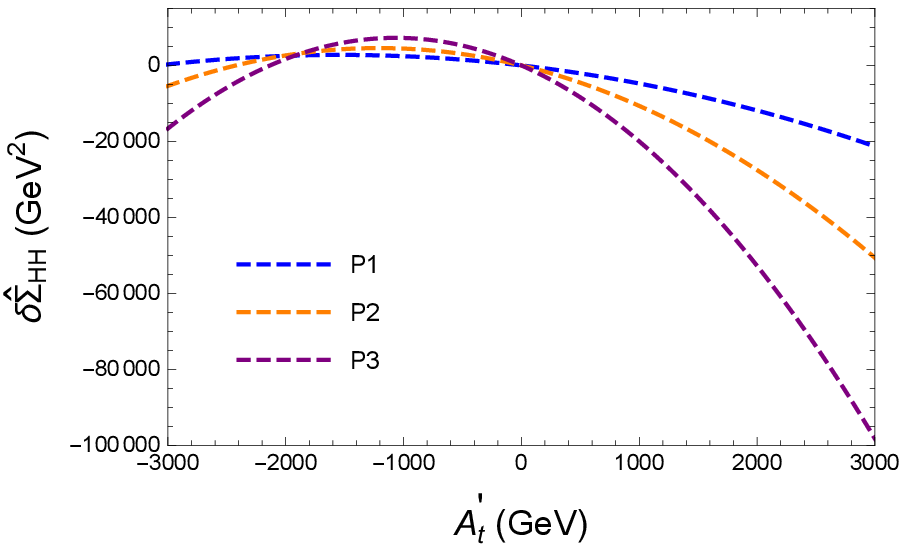  ,scale=0.75,angle=0,clip=}
\hspace{0.5cm}
\psfig{file=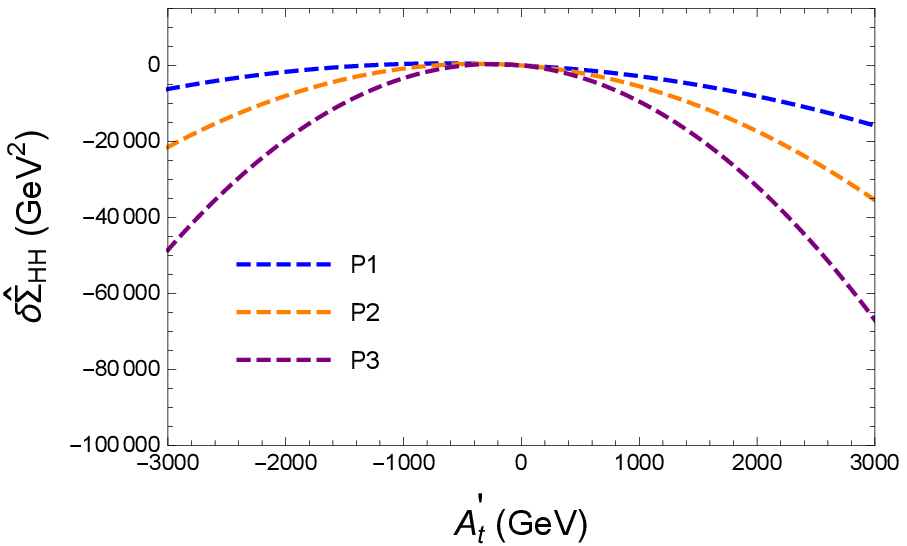 ,scale=0.75,angle=0,clip=}
\end{center}
\caption{ $\delta \hSi_{HH}$  as a funtion of $A^{\prime}_t$ for
  $M_h^{125}$ (left) and 
  $M_h^{125} (\tilde{\chi})$ (right plot).}    
\label{fig:SEHH}
\end{figure}

\begin{figure}[ht!]
\vspace{1em}
\begin{center}
\psfig{file=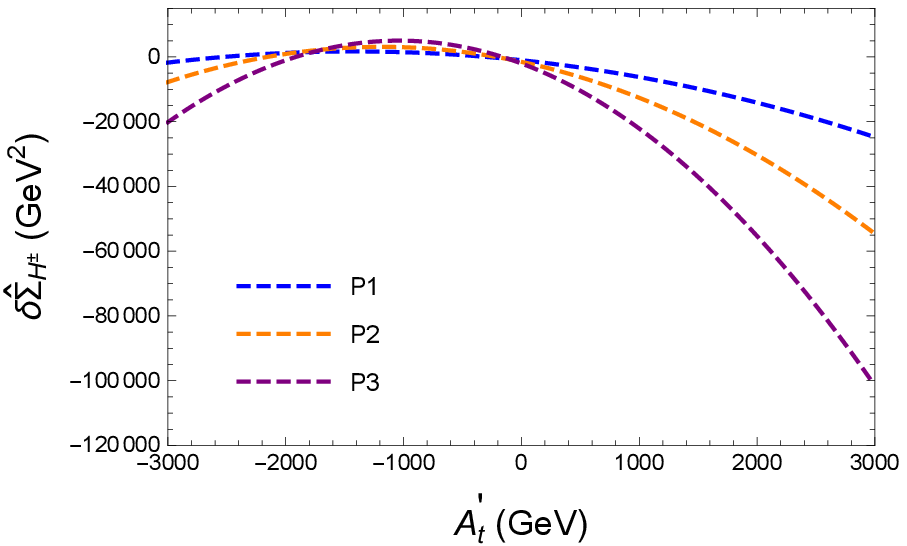  ,scale=0.75,angle=0,clip=}
\hspace{0.5cm}
\psfig{file=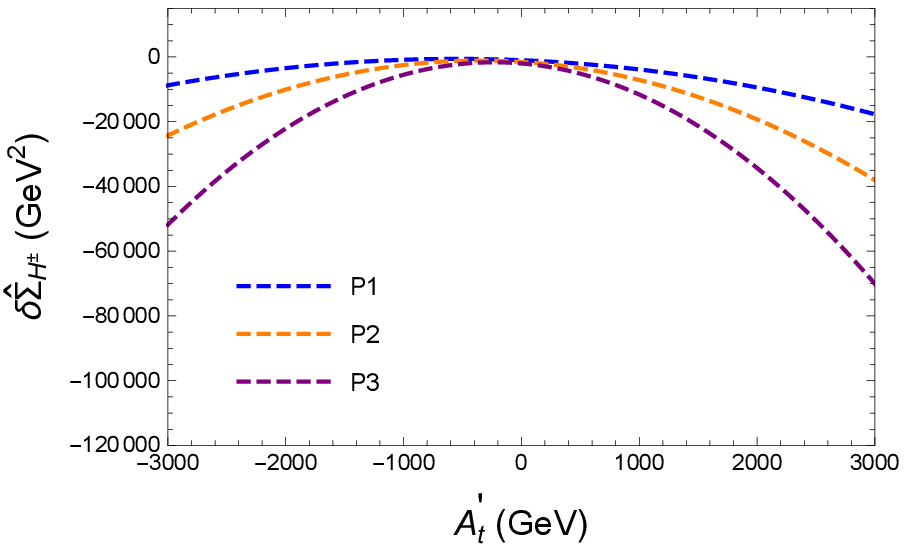 ,scale=0.75,angle=0,clip=}
\end{center}
\caption{ $\delta \hSi_{H^{\pm}}$  as a funtion of $A^{\prime}_t$ for
  $M_h^{125}$ (left) and 
  $M_h^{125} (\tilde{\chi})$ (right plot).}
\label{fig:SEHp}
\end{figure}

The contributions of the non-holomorphic trilinear coupling
$A_t^{\prime}$ to the renormalized Higgs boson self energies,
$\de\hSi_{hh}$, $\de\hSi_{hH}$, $\de\hSi_{HH}$ and $\de\hSi_{H^\pm}$ are shown
as a function of $A_t^\prime$ 
in \reffis{fig:SEh0}, \ref{fig:SEh0HH}, \ref{fig:SEHH} and \ref{fig:SEHp},
respectively. We have varied $A_t^\prime$ in the interval $-3000 \gev$
to $+3000 \gev$. In each figure we show in the left (right)
plot the results for the $M_h^{125}$
($M_h^{125}(\tilde\chi)$) scenario for P1 (P2,
P3) in blue (orange, violet) dashed lines.
The results in the $M_h^{125}(\tilde\tau)$ scenario are effectively
identical to the ones obtained in the $M_h^{125}$ scenario, as could
be expected from the identical parameter values in the scalar top
sector. Consequently, we refrain from showing them separately. On the
other hand, as
can be seen from these figures
the results for the renormalized Higgs-boson self-energies differ
slightly between $M_h^{125}$, $M_h^{125}(\tilde\tau)$ and
$M_h^{125} (\tilde{\chi})$.
This can be traced back to the different $A_t$ values found in
the three scenarios, which in turn stem from the different baseline
values of $X_t$ and in particular $\mu$ in $M_h^{125}$,
$M_h^{125}(\tilde\tau)$ 
w.r.t.\ $M_h^{125}(\tilde\chi)$. As an example, for $\tb = 45$ we
find an intervals of $A_t^{\rm NHSSM} = 2755 \gev$ to $2888 \gev$
for the first two benchmarks, whereas
$A_t^{\rm NHSSM} = 2437 \gev$ to $2570 \gev$ in the latter.

For the renormalized self-energies of the neutral $\cp$-even
Higgs bosons we observe that $\de\hSi_{HH} > \de\hSi_{hH} > \de\hSi_{hh}$.
This can be understood from the fact that the new NH soft SUSY-breaking
term $A_t^\prime$ couples preferrably to the first Higgs doublet, see
\refeq{NonH-TrilinearTerms}. The light \cp-even Higgs, $h$, has a large
contribution from the second Higgs doublet, whereas the $H$ has a large
part from the first Higgs doublet. Consequently, the largest effects are
expected in the coupling of the heavy \cp-even Higgs to scalar tops. The
largest effects on $\hSi_{hh}$ are found in P1 with $\sim \mp 6 \gev^2$
for $A_t^\prime = \mp 3000 \gev$, respectively, with only a small
variation between the three benchmark scenarios. The effect increases to
$\mp 1500 \gev^2$ for $\hSi_{hH}$, nearly equal for all benchmarks and
points. The largest effects are found for $\hSi_{HH}$, reaching up to
$\sim -100000 \gev^2$ for P3 in the $M_h^{125}$ and
$M_h^{125}(\tilde\tau)$ scenario for $A_t^\prime = 3000 \gev$, and up to 
$\sim -70000 \gev^2$ for P3 in the $M_h^{125}(\tilde\chi)$. For
$\hSi_{HH}$ a strong variation between the three $\MA$-$\tb$
combinations can be observed, where larger $\MA$, which in turn allows
for larger $\tb$ leads to the most sizable effect. This can be
understood from the corresponding $\tb$ enhancement of the $A_t^\prime$
contribution. 
Very similar effects can be observed for the renormalized charged
Higgs-boson self-energy, as shown in \reffi{fig:SEHp}. Also for the
charged Higgs-boson, residing largely in the first Higgs doublet, the
$A_t^\prime$ coupling contribution is enhanced with $\tb$,
see \refeq{CHpSqSq}.


\subsection{NH contributions to the Higgs-boson masses}

We now turn to the numerical evaluation of the impact of the NH
trilinear coupling $A_t^\prime$ on the higher-order corrected
Higgs-boson masses themselves. The results shown in the previous
subsection were obtained by subtracting 
the Higgs-boson self-energies values at $A_t^\prime=0$, i.e.\ the pure
MSSM contribution. This allows to directly add these new contributions
to the full calculation of the renormalized Higgs-boson self-energies in
the MSSM. In order to estimate their
effects on the Higgs-bosons masses, we fed these results to the code \fh\
using the \fh\ function {\tt FHAddSelf}. This function adds the NHSSM
contributions to the renormalized Higgs boson self-energies in the
MSSM, evaluated at the highest level of precision.
For details see the discussion at the end of \refse{sec:strategy0}
  and in \refse{sec:strategy}.

\begin{figure}[ht!]
\begin{center}
\psfig{file=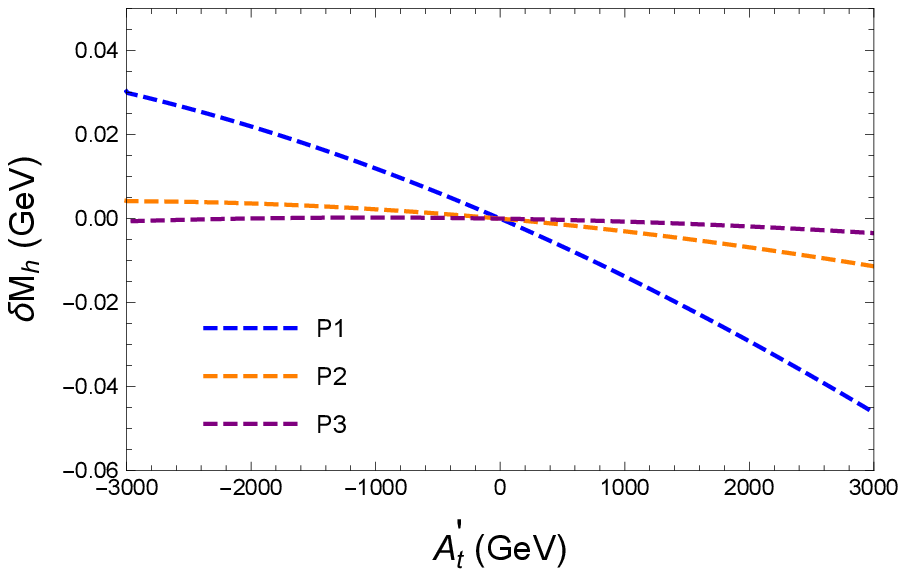  ,scale=0.75,angle=0,clip=}
\hspace{0.5cm}
\psfig{file=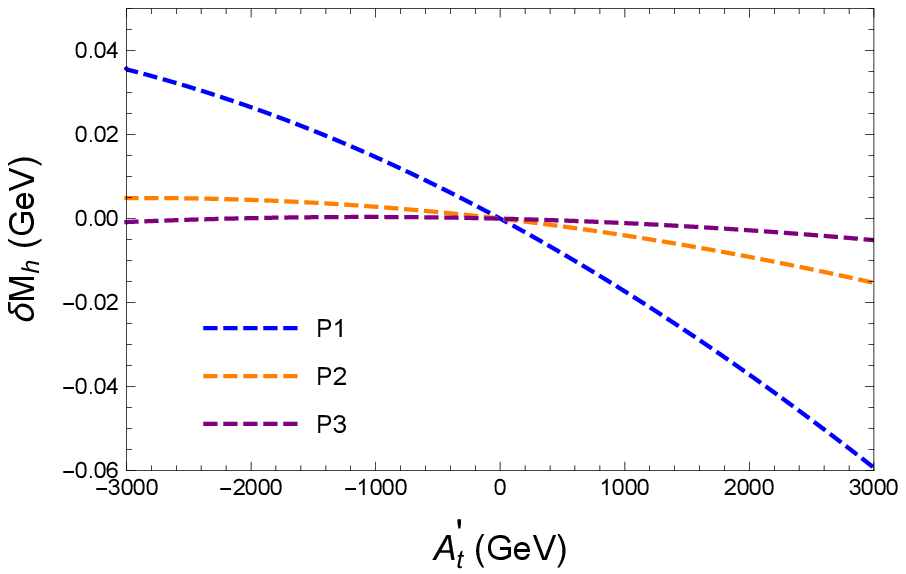 ,scale=0.75,angle=0,clip=}
\end{center}
\caption{ $\delta M_h$  as a funtion of $A^{\prime}_t$ for $M_h^{125}$
  (left) and $M_h^{125}(\tilde{\chi})$ (right plot).
  The results in
    $M_h^{125}(\tilde{\tau})$ are effectively identical to $M_h^{125}$
    and consequently not shown.}
\label{fig:Mh0}
\end{figure}

\begin{figure}[ht!]
\begin{center}
\psfig{file=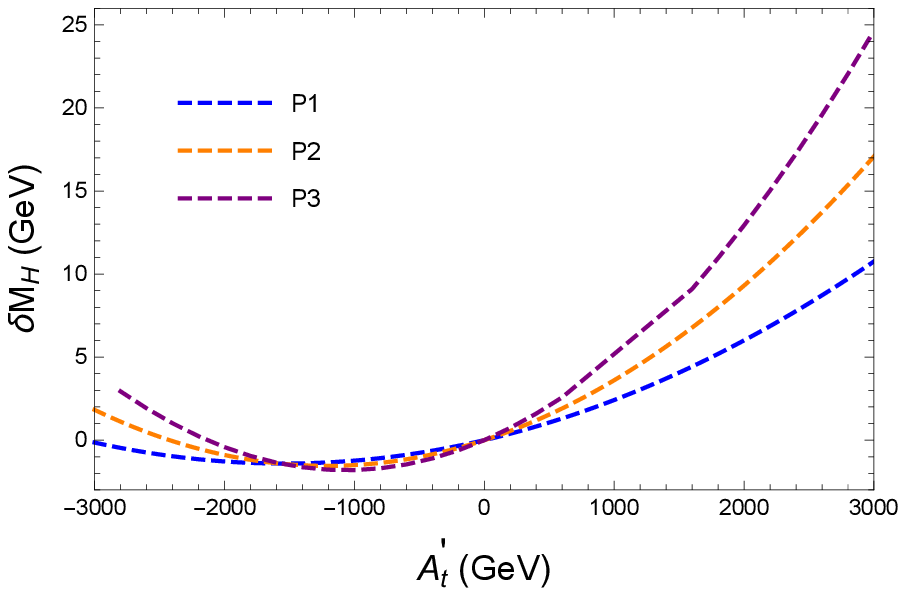  ,scale=0.75,angle=0,clip=}
\hspace{0.5cm}
\psfig{file=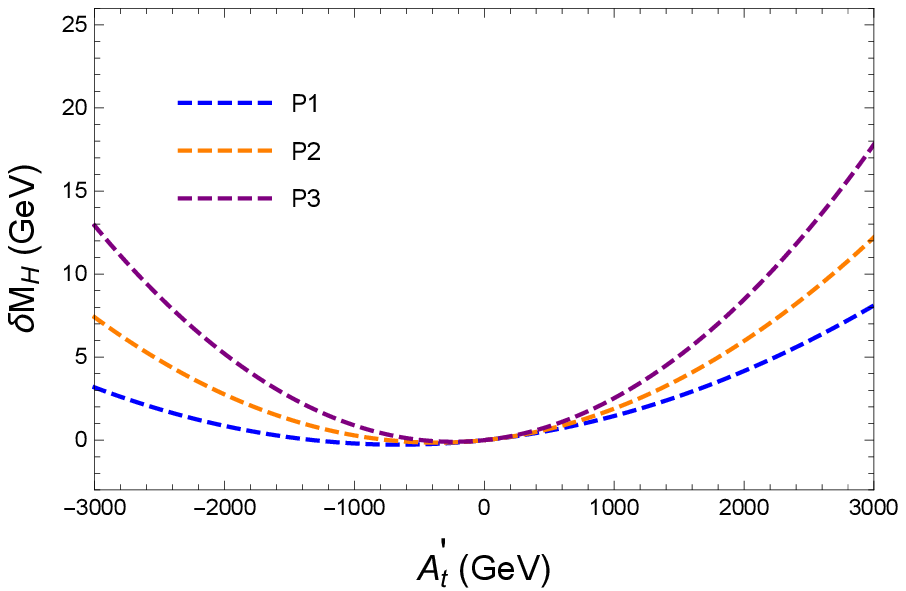 ,scale=0.75,angle=0,clip=}
\end{center}
\caption{ $\delta M_H$  as a funtion of $A^{\prime}_t$ for $M_h^{125}$
  (left), and $M_h^{125}(\tilde{\chi})$ (right plot).}
\label{fig:MHH}
\end{figure}

\begin{figure}[ht!]
\begin{center}
\psfig{file=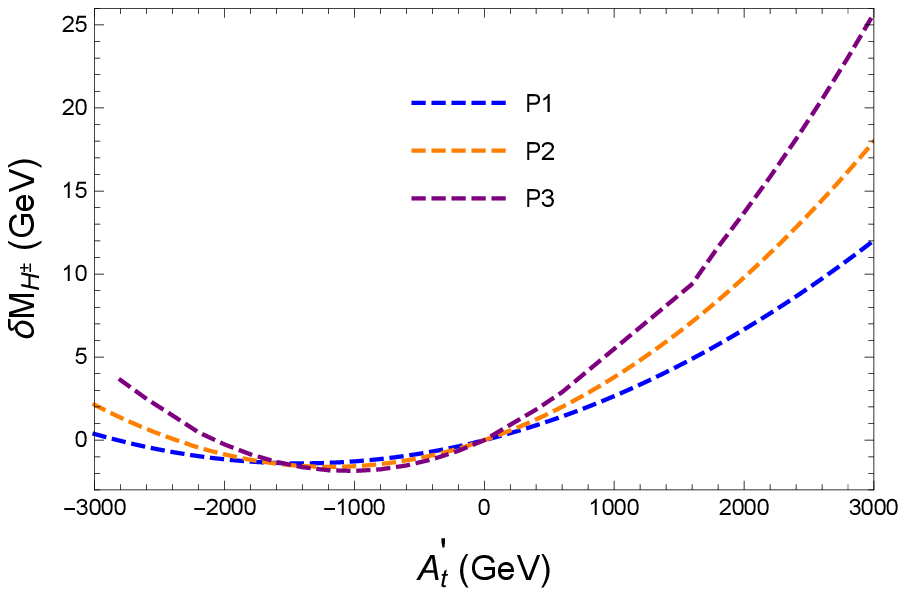  ,scale=0.75,angle=0,clip=}
\hspace{0.5cm}
\psfig{file=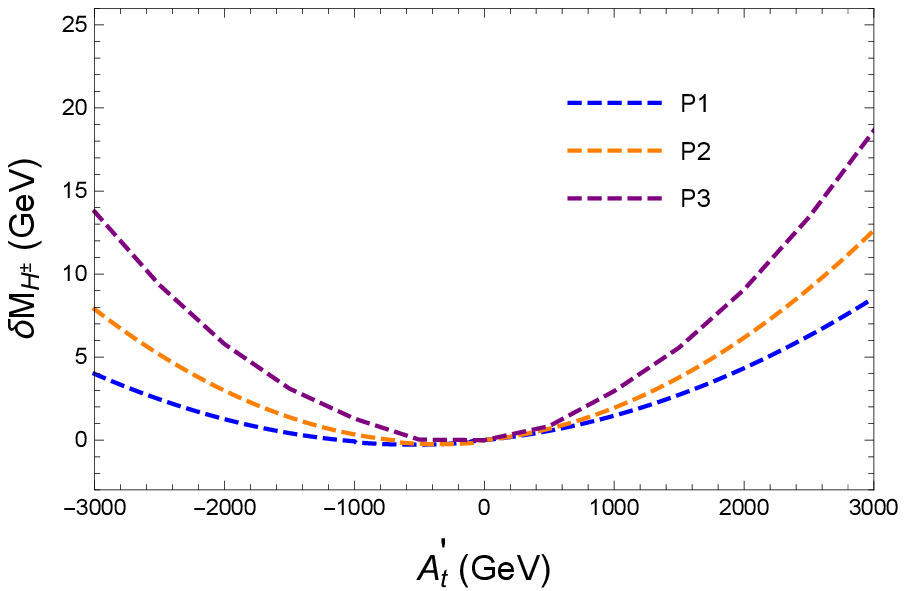 ,scale=0.75,angle=0,clip=}
\end{center}
\caption{ $\delta M_{H^{\pm}}$  as a funtion of $A^{\prime}_t$ for
  $M_h^{125}$ (left) and $M_h^{125} (\tilde{\chi})$ (right plot).}
\label{fig:MHp}
\end{figure}

The obtained results are shown as a function of $A_t^\prime$ in
\reffis{fig:Mh0}, \ref{fig:MHH} and \ref{fig:MHp} for $\de\Mh$, $\de\MH$ and
$\de\MHp$, respectively. As in the previous subsection, we use the interval
of $A_t^\prime = -3000 \gev$ to $+3000 \gev$.
The order of the plots and the color coding is as in the previous subsection.
In particular, we again do not show the resuls for
$M_h^{125}(\tilde\tau)$, as they are effectively identical to the ones
in the $M_h^{125}$ scenario.
Since the effects on the
renormalized Higgs-boson self-energies follows the pattern
$\de\hSi_{HH} > \de\hSi_{hH} > \de\hSi_{hh} \sim \de\hSi_{H^\pm}$,
one expects larger effects for the two heavy Higgs-boson masses than for
the light \cp-even Higgs. Only for very large values of
$\MA^2 \gg |\de\hSi_{HH}|, |\de\hSi_{H^\pm}|$ the additional
contributions from NH terms should become irrelevant for $\MH$ and
$\MHp$. 

For $\de\Mh$, as shown in \reffi{fig:Mh0}, the NH contributions
yield corrections are in general found to be very small, as could
be expected from the size of $\de\hSi_{hh}$, see \reffi{fig:SEh0}.
They reach up to $\sim -45 \mev \gev$ for $A_t^\prime = 3000 \gev$
in the $M_h^{125}$ and $M_h^{125}(\tilde\tau)$ scenario for P1, with
negligible changes in P2 and P3.
In these two benchmark
scenarios the corrections for negative $A_t^\prime$ stays below $+30 \mev$.
In the $M_h^{125}(\tilde\chi)$ scenario
the results look similar, with slightly larger corrections in P1.
The fact that P1 exhibits the largest corrections
corroborates that this effect on $\Mh$, as expected, stems 
from the contribution of $\de\hSi_{hh}$. 
Since the corrections turn out to be very small over the whole
analyzed parameter space demonstrates that the NH terms {\em do not}
alleviate the fact that large stop masses are needed to reach the value
of $\Mh \sim 125 \gev$. On the other hand, effects from
$\Ab^\prime$ and/or $\Atau^\prime$ could show a different behavior. We
leave this analysis for future work.
It should be noted here that the size of the numerical effects on
$\Mh$ found here are substantially smaller than previously claimed in
the literature~\cite{Chattopadhyay:2016ivr}. This can be explained by
the fact that we ensured to compare results including the NH effects to
the ``pure MSSM'', but leaving the physics scenario (the stop masses and
mixing) unchanged.

The changes in the heavy $\cp$-even Higgs-boson mass, $\MH$, are shown in
\reffi{fig:MHH}. The general pattern follows the size of the corrections
for $\Mh$, as analyzed in \reffi{fig:Mh0}. However, for $\MH$ the
corrections turn out to be in general positive. The largest values
reached are $\sim +25 \gev$ for $A_t^\prime$ in P3 in the $M_h^{125}$ and the
$M_h^{125}(\tilde\tau)$ scenario. In the $M_h^{125}(\tilde\chi)$
scenario the largest corresponding value is $\sim +18 \gev$. For
$A_t^\prime = -3000 \gev$ the corrections reach up to $+5 \gev$ in P3
for the first two benchmarks, and up to $+13 \gev$ for the third, with
correspondingly smaller values for P2 and P1.

As a last step, we show the changes in the charged Higgs boson mass,
$\MHp$, in \reffi{fig:MHp}. 
As can be expected from the NH contributions to the renormalized
Higgs-boson self-energies, which are similar for $\de\hSi_{HH}$ and
$\de\hSi_{H^\pm}$, see \reffis{fig:SEHH} and \ref{fig:SEHp}, also the
correction to the two heavy Higgs-boson masses themselves turn out to
be similar. $\de\MHp$ follows in sign and size the corrections found
for $\MH$. The NH contributions {\em do not} lead to an enhanced
splitting between the $\MH$ and $\MHp$, but only to larger differences
between $\MA$ (our input) and the other two heavy Higgs-boson masses.

\section{Conclusions}
\label{sec:conclusions}

In this paper we have investigated the effect of non-holomorphic soft
SUSY-breaking terms to the Higgs-boson mass preditions in the MSSM, a
model dubbed NHSSM. In order to perform the calculations 
we generated the \fa\ model file using mathematica package SARAH. The
model file was then used in the \fa/\fc\ setup (including 
modifications in the \fc\ driver files to addapt the NHSSM specific
input) to generate analytical and
numerical results for the various renormalized Higgs boson self
energies. We concentrated on the contributions from the top/scalar top
sector. The relevant NH term is the trilinear coupling $A_t^\prime$.
The results for the renormalized Higgs-boson self-energies were then fed
into the code \fh\ (using the 
{\tt FHAddSelf} subroutine) to calculate the predictions for the Higgs
boson masses.

We took particular care to analyze the pure NH contribution. The
$A_t^\prime$ contributions enter into the scalar top mass matrix via the
non-diagonal entry $X_t$, as
well as into the Higgs-stop couplings. An analsysis simply varying
$A_t^\prime$ thus leads to a shift in the scalar top
masses, which should be considered a different physics scenario,
as the stop masses and mixing angle are expected to be measured in
the future (if SUSY is realized).
Consequently, an observed effect from a naive variation of
$\At^\prime$ can possibly be
mimicked by a change in the holomorphic soft SUSY-breaking terms, in
particular the trilinear Higgs-stop coupling, $A_t$: for each choice of
$A_t^\prime$ the parameter $A_t$ can be adjusted to yield the same
scalar top mass. An observed scalar top mass spectrum thus corresponds
to a continuous set of combinations of $A_t$ and $A_t^\prime$ (keeping
the other soft SUSY-breaking parameters and $\mu$ fix). An analysis that
simply varies $A_t^\prime$, resulting in shifts in the scalar top
masses, can thus not be regarded realistic. Therefore, in our analysis
we required $X_t$ to be constant under a change of $A_t^\prime$ by an
adjustment of $A_t$. In this way the effect of the NH contributions is
shifted into the Higgs-stop couplings and can readily be analyzed.

For the NH contributions to the renormalized Higgs-boson self-energies
we find $\de\hSi_{hh} < \de\hSi_{hH} < \de\hSi_{HH} \sim \de\hSi_{H^\pm}$.
This can be understood from the fact that the new NH soft SUSY-breaking
term $A_t^\prime$ couples preferrably to the first Higgs doublet.
The light \cp-even Higgs,~$h$, has a large
contribution from the second Higgs doublet, whereas the $H$, as well as
the charged Higgs have their largest component
from the first Higgs doublet. Consequently, the largest effects are
expected in the coupling of the heavy \cp-even Higgs, or the charged
Higgs to scalar tops.

For the numerical analysis we chose three LHC benchmark scenarios
($M_h^{125}$, $M_h^{125}(\tilde\tau)$ and $M_h^{125}(\tilde\chi)$),
and in each scenario three combinations of $(\MA, \tb)$ that are allowed
by current MSSM Higgs-boson searches at the LHC,
$(1000 \gev, 7), (1500 \gev, 15), (2000 \gev, 45)$, called P1, P2, P3,
respectively. 
$A_t^\prime$ has been varied from $-3000 \gev$ to $+3000 \gev$.
The results in the $M_h^{125}$ and the $M_h^{125}(\tilde\tau)$
scenario are effectively identical due to their identical settings in
the scalar top sector. The results in the $M_h^{125}(\tilde\chi)$
scenario, however, can vary substantially from the other two scenarios.
For $\de\Mh$ the NH contributions
yield corrections are in general found to be very small, in
contrary to previous claims in the literature.
They reach up to $\sim -60 \mev$ in the analyzed parameter space, where
P1 exhibits the largest corrections.
Since the corrections turn out to be very small over the whole
analyzed parameter space we find that the NH terms {\em do not}
alleviate the fact that large stop masses are needed to reach the value
of $\Mh \sim 125 \gev$. 
The situation might change for the corrections involving
$\Ab^\prime$ and/or $\Atau^\prime$, which we leave for future work.
The numerical effects for $\MH$ and $\MHp$ were found to be in general
positive and reached values of up to $+25 \gev$ for $\MH$ and $\MHp$.

Despite the fact that the NH contributions entering via $A_t^\prime$ are
small for $\Mh$, a full analysis of supersymmetric extensions
of the SM should 
include the possibility of NH contributions. We aim for an inclusion
of these effects into the code \fh.

\subsection*{Acknowledgments}

We thank F.~Staub for helpful discussions on SARAH
and the model file generation for \fa.  
The work of S.H.\ has received financial support from the
grant PID2019-110058GB-C21 funded by
MCIN/AEI/10.13039/501100011033 and by ``ERDF A way of making Europe".
MEINCOP Spain under contract PID2019-110058GB-C21 and in part by
by the grant IFT Centro de Excelencia Severo Ochoa CEX2020-001007-S
funded by MCIN/AEI/10.13039/ 501100011033.



\end{document}